\documentclass[reprint, 
superscriptaddress,
 amsmath,amssymb,
 aps, 
pra,
floatfix
]{revtex4-2} 
\usepackage{xfrac}
\usepackage{lmodern}
\usepackage{graphicx}
\usepackage{dcolumn}
\usepackage{bm}
\usepackage{hyperref}

\hypersetup{
	colorlinks = true,
	urlcolor   = blue,
	linkcolor  = blue,
	citecolor  = blue
}

\newcommand*{\ket}[1]{| #1 \rangle} 

\long\def\comment#1{}

\usepackage[T1]{fontenc}
\usepackage[utf8x]{inputenc}
\usepackage{graphicx}
\usepackage[dvipsnames]{xcolor}
\usepackage{upgreek}
\usepackage{hhline}
\begin{document}
\preprint{APS/123-QED}
\title{Implementation of two-qubit Rydberg operations on neutral Rb-87 atoms in systems with different intermediate states}

\author{I. V. Iukhnovets}
\email{Contact author: i.yukhnovets@rqc.ru}
\affiliation{Moscow Institute of Physics and Technology (MIPT), 141700 Dolgoprudny, Moscow Region, Russia}
\affiliation{Russian Quantum Center (RQC), 143025 Skolkovo, Russia}
\affiliation{P. N. Lebedev Physical Institute (LPI), 119991 Moscow, Russia}
\author{O.~V.~Bychkova}%
\affiliation{P. N. Lebedev Physical Institute (LPI), 119991 Moscow, Russia}
\author{I. B. Bobrov}%
\affiliation{Quantum Technology Centre and Faculty of Physics, M. V. Lomonosov Moscow State University, 119991 Moscow,
Russia}
\author{A. P. Gordeev} 
\affiliation{P. N. Lebedev Physical Institute (LPI), 119991 Moscow, Russia}
\affiliation{Quantum Technology Centre and Faculty of Physics, M. V. Lomonosov Moscow State University, 119991 Moscow,
Russia}
\author{M. Y. Goloshchapov}
\affiliation{Technical University of Munich, Arcisstraße 21, 80333, Munich, Germany}
\affiliation{Ludwig-Maximilians-Universität München, Geschwister-Scholl-Platz 1, 80539, Munich, Germany}
\author{G. I. Struchalin}%
\affiliation{Quantum Technology Centre and Faculty of Physics, M. V. Lomonosov Moscow State University, 119991 Moscow,
Russia}
\author{S. S. Straupe}%
\affiliation{Russian Quantum Center (RQC), 143025 Skolkovo, Russia}
\affiliation{Quantum Technology Centre and Faculty of Physics, M. V. Lomonosov Moscow State University, 119991 Moscow,
Russia}

\date{\today}

\begin{abstract}
This work presents an experimental setup for implementing two-qubit operations on neutral atoms ($^{87}$Rb) with the possibility of using two different Rydberg excitation schemes. One of them uses 5P$_{1/2}$ as the intermediate level and applies the second-stage beam locally to the addressed atoms. The second scheme uses the 6P$_{3/2}$ level; in this scheme, the particles to be entangled are moved to a separate zone through which both Rydberg beams pass. The advantages and limitations of both schemes are analyzed. Based on numerical modeling performed with a Julia package developed by the authors, it is demonstrated that the spatial configuration has a greater effect on quantum-operation fidelity than the choice of intermediate level. An experimental implementation of the scheme using the 6P$_{3/2}$ level is demonstrated, making it possible to achieve a two-qubit operation fidelity of 94\%.
\end{abstract}

\maketitle

\section{Introduction}
For the practical implementation of universal quantum computation, at least three properties must be combined: long coherence time, a high-fidelity set of basic logical operations, and an architecture that allows scaling the number of qubits without fundamentally complicating control \cite{Preskill:2018:2}. The third property is especially characteristic of neutral-atom platforms, which distinguishes them from other platforms (ion, superconducting, etc.). In addition, universal quantum computation requires at least one two-qubit entangling operation \cite{NielsenChuang2010}. \par
Among the main advantages of atomic platforms, in addition to the aforementioned scalability, are the natural identity of atoms of the same isotope, which in the absence of inhomogeneous external (stray) electromagnetic fields eliminates the need for individual calibration, and the encoding of qubits in hyperfine sublevels of the ground state, which provides long coherence times for qubit states \cite{Saffman:2010:82}. \par
The $^{87}$Rb isotope is one of the most widely used elements in this field because of several factors: its well-studied atomic structure, the presence of convenient closed cycling transitions for laser cooling on the $D_2$ line, the availability of laser sources with substantial power at the required wavelengths, and a favorable structure of hyperfine sublevels, which helps increase the coherence time. \par
In recent years, technical capabilities have made it possible to obtain arrays of several hundred computational qubits represented by rubidium atoms \cite{Ebadi:2021:595,Bluvstein:2026:649}. Ref.~\cite{Manetsch:2025:647} presents an array of cesium atoms consisting of 6100 qubits, but it is important to note that two-qubit operations are not performed on them; only the capability of coherent control is demonstrated. \par
In most experimental setups of the class considered here, two-qubit gates are implemented based on the Rydberg blockade effect: excitation of one atom to a Rydberg state shifts the energy levels of a neighboring atom located within the blockade radius, making excitation of the second atom to the same state impossible \cite{Saffman:2010:82,Jaksch:2000:85,Lukin:2001:87,Walker:2012:61}. This phenomenon makes it possible to implement controlled two-qubit operations of the CZ and CNOT type. \par
Whereas a fidelity of 99.997\% has been achieved for single-qubit operations \cite{Sheng:2018:121,Bluvstein:2024:626}, two-qubit operations are currently characterized by error values that are 2--3 orders of magnitude higher \cite{Bluvstein:2026:649, Evered:2023:622}. \par
\begin{figure}[t]
\centering\includegraphics[height=0.67\linewidth]{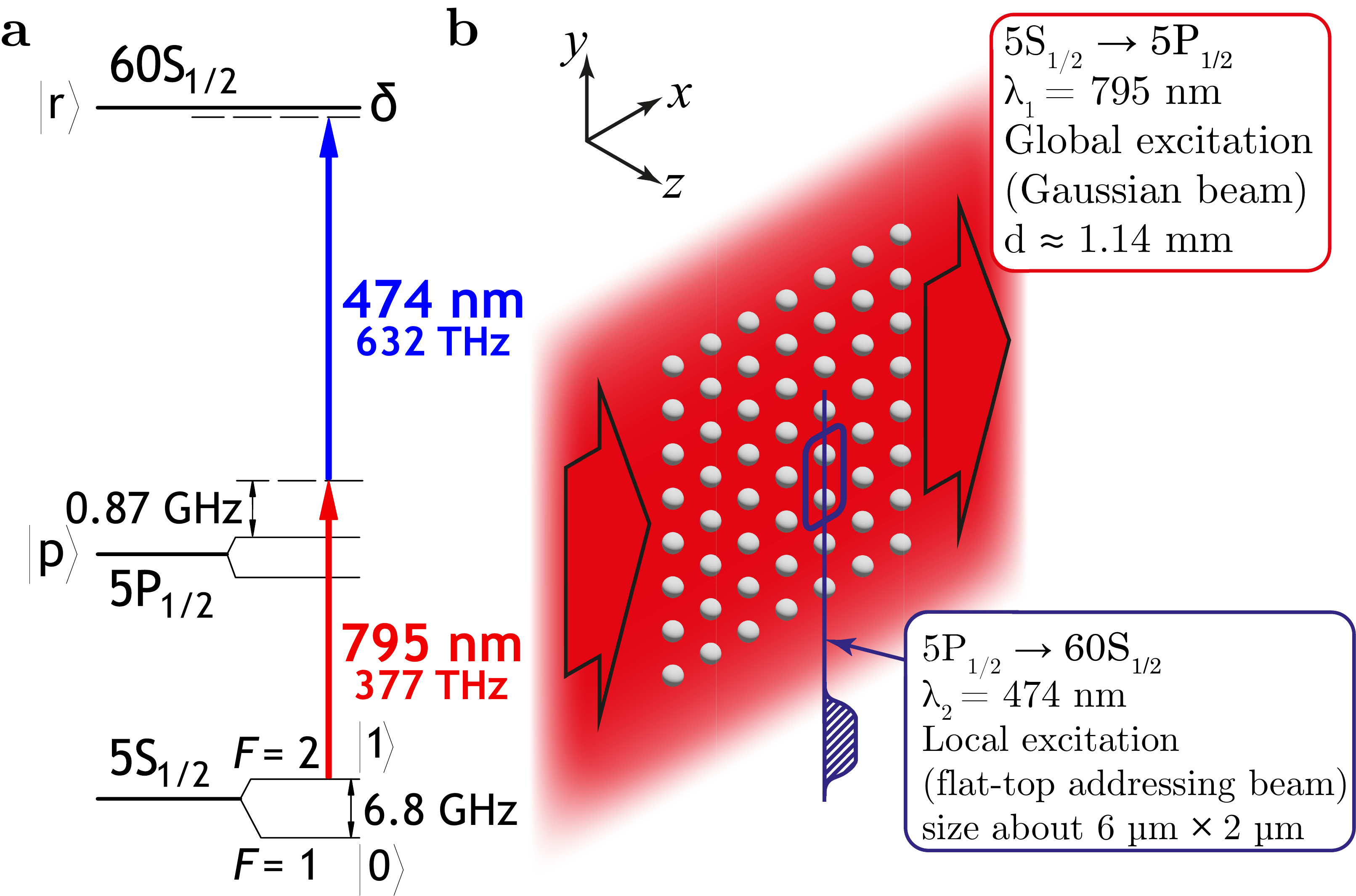}
\caption{Scheme for implementing excitation through the intermediate level 5P$_{1/2}$. (a) Scheme of two-photon excitation to the state $\ket{r}$ through the intermediate state $\ket{p}$. (b) Experimental implementation. The beam for the first excitation stage is global and propagates along the $OX$ direction. The beam for the second excitation stage provides site-selective addressing, propagates along $OZ$, and has a flat-top intensity profile}\label{fig:scheme5p}
\end{figure}
\begin{figure}[t]
\centering\includegraphics[height=0.67\linewidth]{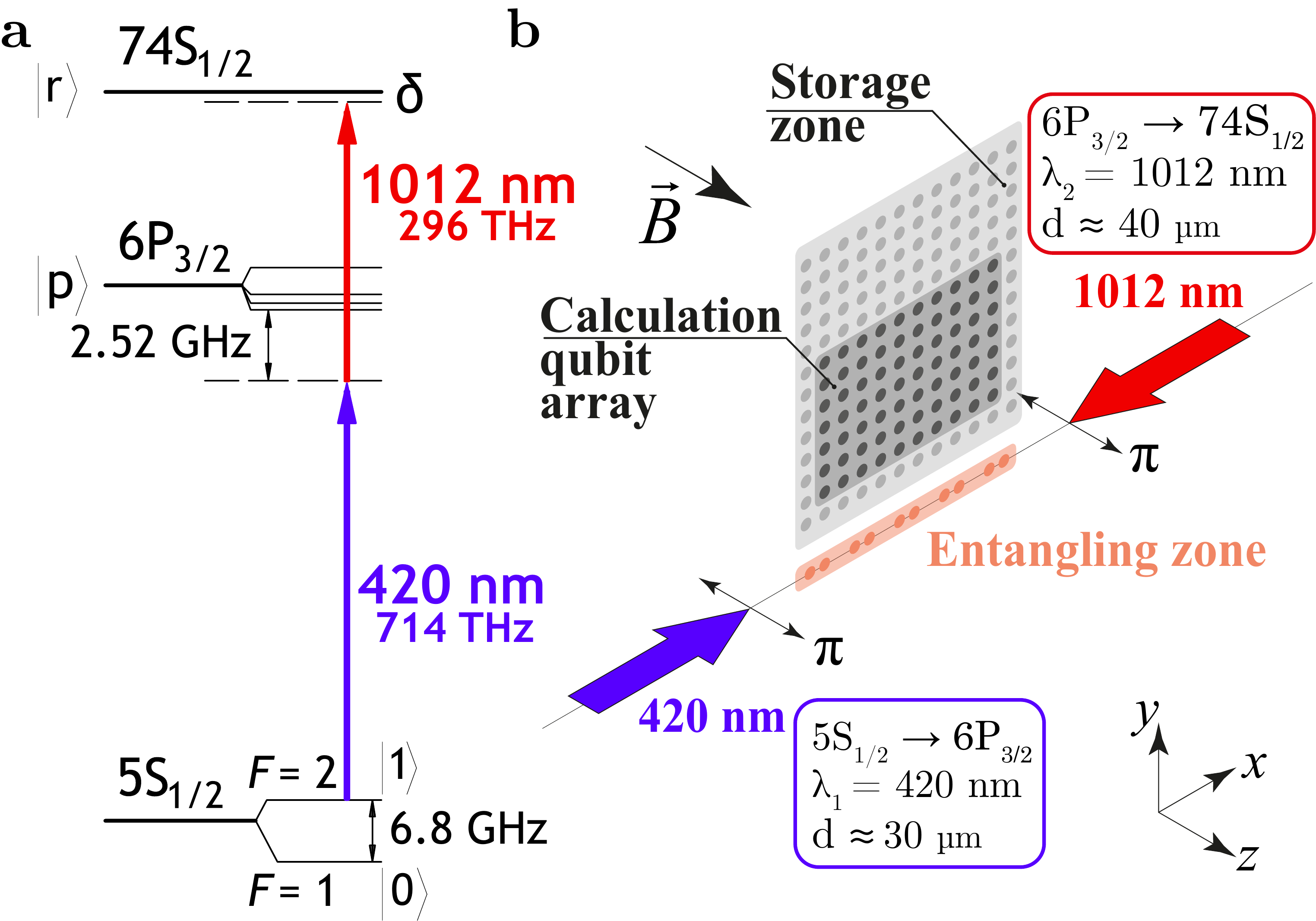}
\caption{Scheme for implementing excitation through the intermediate level 6P$_{3/2}$. (a) Scheme of two-photon excitation to a Rydberg state. (b) Experimental implementation. Both beams act on atoms in the entanglement zone. The first-stage beam propagates along the $OX$ axis, and the second-stage beam propagates in the opposite direction}\label{fig:scheme6p}
\end{figure}

In this work, we consider schemes of two-photon Rydberg excitation both through the intermediate level 5P$_{1/2}$ (shown in Fig.~\ref{fig:scheme5p}a; below, for brevity, we will call it ``scheme A'') and through the 6P$_{3/2}$ level (Fig.~\ref{fig:scheme6p}a; below, ``scheme B''). The implementations of these schemes differ in their spatial configuration: in the system with 5P$_{1/2}$, the second-stage beam selectively addresses the particles to be entangled, whereas in the system with 6P$_{3/2}$, zone separation is used: the atoms to be entangled are transferred to a special zone through which the Rydberg excitation beams pass. Before describing our experiment, we will briefly review similar systems, since both the first and the second states are actively used as intermediate states in setups of this type.

In Refs.~\cite{Levine:2018:121,Levine:2019:123}, individual $^{87}$Rb atoms were moved to an entanglement zone (which also served as the computation zone), a one-dimensional array of optical tweezers through which the beams of both Rydberg excitation stages passed. The 6P$_{3/2}$ intermediate level was used. Weakly damped Rabi oscillations of Rydberg excitation and entangled states with a fidelity of 0.974 were demonstrated, which became one of the key early milestones in the field considered here. \par
In Ref.~\cite{Bluvstein:2022:604}, a two-dimensional reconfigurable system was implemented, in which hyperfine states were used for storing and transporting quantum information, while Rydberg excitation was used to perform two-qubit operations. In 2026, the same laboratory created a zoned architecture with elements of fault-tolerant computation on arrays of up to 448 $^{87}$Rb atoms \cite{Bluvstein:2026:649}. At present, the Rydberg beams passing through the entanglement zone in this setup have a ``flat-top'' profile (a beam with a flat intensity and phase profile). \par
The intermediate levels 5P$_{1/2}$ and 5P$_{3/2}$ and local excitation are also used \cite{deLeseleuc:2018:97,Fu:2022:105,Chew:2022:16,Li:2025:16}. Ref.~\cite{Li:2025:16} proposed and experimentally demonstrated a fiber architecture for quantum computation on $^{87}$Rb. In this system, Rydberg blockade between neighboring atoms was shown under two-photon excitation through the 5P$_{3/2}$ level to the 68D$_{5/2}$ state.

In Ref.~\cite{Miles:2026:arxiv}, in a two-element Rb--Cs system, rubidium atoms are locally excited through the 6P$_{1/2}$ level. \par
Thus, static architectures with built-in individual addressing are also a promising direction, primarily because there is no need to move atoms between zones. This gives two advantages at once: the ability not to spend resources on moving atoms and a reduction of the gate implementation time. \par
In our work, we present for the first time an experimental setup capable of implementing both variants of Rydberg excitation. Its creation is a prerequisite for improving both methods on the same quantum-computer prototype. 

\section{Theoretical Foundations \label{sec:theory}}
The excitation of a $^{87}$Rb atom from the ground state $\ket{g} =$ 5S$_{1/2}$ to the Rydberg state $\ket{r_n}=n$S$_{1/2}$ ($n \sim 50 \dots 80$) is a two-photon process carried out through a virtual intermediate state. To minimize spontaneous scattering from the intermediate state, a detuning on the order of several GHz is introduced. \par
The two-photon Rabi frequency for excitation $|g\rangle \to |r\rangle$ through a virtual intermediate level $|p\rangle$ is expressed as
\begin{equation}
\Omega_\text{2ph} = \frac{\Omega_1 \Omega_2}{2\Delta},
\end{equation}
where $\Omega_1$ and $\Omega_2$ are the Rabi frequencies of the first- and second-stage transitions, respectively, and $\Delta$ is the detuning from the intermediate state \cite{Saffman:2010:82}. The Rabi frequency is defined as $\Omega = d \cdot \mathcal{E}/\hbar$, where $d$ is the dipole matrix element and $\mathcal{E}$ is the electric-field amplitude of the laser radiation. \par
The matrix elements of transitions to highly excited Rydberg states are relatively small, which requires the use of high intensities for the second-stage beams. \par
Spontaneous photon scattering from the intermediate level is the main source of decoherence in two-photon Rydberg gates; its rate is determined by the expression \cite{Saffman:2010:82, Walker:2012:61}
\begin{equation}
R_\text{sc} \propto \frac{\Omega_1^2 \Gamma_e}{4\Delta^2},
\end{equation}
where $\Gamma_e$ is the natural linewidth of the intermediate level. Thus, to suppress scattering it is necessary to increase the detuning $\Delta$ while simultaneously maintaining a sufficient two-photon Rabi frequency. \par
For the 5S$_{1/2} \to \ $5P$_{1/2}$ transition (the D$_1$ line, 795~nm), the reduced dipole matrix element is $\langle 5\mathrm{S}_{1/2} | er | 5\mathrm{P}_{1/2} \rangle \approx 4.23\, ea_0$ \cite{Steck_D1}. The radiative lifetime of the $5\mathrm{P}_{1/2}$ state is $\tau_{5\mathrm{P}_{1/2}} = 27.75(8)$ ns, which corresponds to a natural linewidth $\Gamma_{5\mathrm{P}}/2\pi \approx 5.75$ MHz. \par
For the 5S$_{1/2} \to \ $6P$_{3/2}$ transition (420 nm), the experimentally determined dipole matrix element is $0.5230(8)\,ea_0$. The lifetime of the 6P$_{3/2}$ level is $\tau_{6\mathrm{P}_{3/2}} \approx 112$ ns \cite{Safronova:2011:83, Gomez:2004:21}, giving a natural linewidth $\Gamma_{6\mathrm{P}}/2\pi \approx 1.4$~MHz, approximately four times smaller than for the 5P$_{1/2}$ level. Thus, at the same detuning $\Delta$, the scattering rate from the 6P$_{3/2}$ level is approximately four times lower than from 5P$_{1/2}$. The first-stage dipole matrix element for 5S $\to$ 6P is approximately 8 times smaller than for 5S $\to$ 5P, so an increased first-stage radiation intensity is required. However, this does not pose engineering difficulties, since the first-stage excitation beam at a detuning from the intermediate level on the order of several GHz must have an intensity on the order of W/mm$^2$. \par
Equality of the Rabi frequencies of the two stages makes it possible to eliminate the differential light shift of the ground and Rydberg states. In the experiment, however, exact equality does not occur because of several factors associated with imperfections in the experimental setup. Nevertheless, to avoid a significant increase in the light shift, when increasing the detuning to reduce spontaneous scattering from the level $\ket{p}$, one must increase $\Omega_1$ and $\Omega_2$ in equal proportions. In contrast to the comparatively small required intensity of the first-stage beam, the second-stage beam must reach values on the order of kW/mm$^2$. Technologically, obtaining such a high value at a wavelength of 1012 nm is much easier than at a wavelength of 474~nm, and this is an engineering advantage of scheme B. \par
In Ref.~\cite{Evered:2023:622}, substantial detuning from the intermediate state ($\Delta/2\pi = 7.8$ GHz) while maintaining a high two-photon Rabi frequency ($\Omega/2\pi = 4.6$ MHz) was achieved through excitation to a low-lying Rydberg state ($n = 53$) and an unprecedentedly high (up to 50~W) power of the second-stage laser. In our work, because of the limited tuning range of both lasers and the maximum 1012 nm laser power of 6 W, the $n=74$ level was used. \par
As a minor drawback of scheme B, one can mention the increased residual Doppler sensitivity. For a two-photon transition, it is determined by the sum of the wave vectors of the two laser beams: $\delta\omega_\text{Doppler} = \vec v \cdot \vec{k}_{eff}$, where $v$ is the atom velocity and $\vec{k}_{eff}$ is the effective wave vector. When counterpropagating beams are used, the expression $k_\text{eff} = k_2 - k_1$ holds for the projections. In scheme B, the ratio $k_1/k_2 = \lambda_2/\lambda_1 \approx 2.4$, which is much larger than in scheme A, and therefore the compensation is less effective. This leads to an increased Doppler width of the two-photon transition; however, for cold atoms in optical tweezers the rms thermal velocity ($v_\text{rms} = \sqrt{k_B T / m}$) is 8~cm/s at an atom temperature of 70~$\mu$K (corresponding to our setup), so Doppler broadening is not a limiting factor for gate fidelity. \par
Another disadvantage of scheme B, when targeting fidelities above 99\%, may be heating of the atoms during their transport. Moving an atom is accompanied by its acceleration and deceleration in a movable optical tweezer, which inevitably leads to heating. However, a recent work \cite{Evered:2026:arxiv} demonstrates a raw (i.e., without postselection) two-qubit operation fidelity of 99.854(4)\% using a scheme involving movement and states that the contribution from thermal motion is very small and that the gate infidelity arises mostly of other sources. \par
Thus, the combination of the listed factors makes scheme B with zone separation preferable for implementing high-fidelity quantum gates, which is consistent both with the experimental results of leading groups worldwide \cite{Evered:2023:622, Bluvstein:2024:626, Bluvstein:2026:649} and with the results of our numerical modeling presented in Sec.~\ref{sec:modeling}. \par
The most common method for implementing a CZ gate on Rydberg atoms at present is the Levine--Pichler method, based on geometric phases accumulated during Rabi oscillations between qubit and Rydberg states along closed trajectories on the Bloch sphere \cite{Levine:2019:123}. It will be presented in more detail below in Sec.~\ref{sec:setup} as part of the description of the calibration of the experimental setup parameters. \par
A factor that substantially affects the fidelity of a two-qubit operation is the atom temperature. Motion of atoms due to finite temperature or photon recoil leads to a stochastic accumulation of the gate phase, while increased delocalization of atoms leads to fluctuations of the Rydberg interaction and excitation errors \cite{Robicheaux:2021:103}. The latter effect can be partially suppressed by using a flat-top beam  profile \cite{Evered:2023:622,Gillen:2016:94}.

\section{Numerical Model \label{sec:modeling}}
Because a direct comparison of excitation schemes through the 5P and 6P levels under completely identical geometric conditions, as well as a comparison of two spatial configurations using the same intermediate level, is not experimentally possible, numerical modeling was performed for a quantitative comparison of operation fidelities using our NeutralAtoms.jl library, developed in the Julia programming language \cite{NeutralAtoms}. \par
The library is based on reducing the system to an effective model in which an effective level $\ket{L}$ is introduced, following Ref.~\cite{deLeseleuc:2018:97}. This level effectively represents spontaneous photon scattering from the intermediate level and thereby makes it possible to account for the dominant loss mechanisms without sophistication of management. The dynamics of the effective system are modeled by solving the time-dependent equation in Lindblad form using QuantumOptics.jl \cite{QuantumOptics}. A more detailed description of the NeutralAtoms.jl library is presented in Ref.~\cite{Iukhnovets:2026:arxiv}. \par
Table~\ref{tab:fidelities} gives the fidelity values of Rydberg two-qubit operations as a function of the spatial configuration of the system, the intermediate level $\ket{p}$, and the Rydberg level $\ket{r}$. The parameters of the Rydberg excitation system correspond to those described below in Sec.~\ref{sec:setup}. The obtained results show that the spatial configuration makes the largest contribution to the gate fidelity.
\begin{table}[h]
    \centering
    \caption{Calculated dependence of two-qubit operation fidelity on the intermediate level, geometric configuration, and Rydberg-level number $n_r$ according to modeling results obtained using the NeutralAtoms.jl library}
    \label{tab:fidelities}
    \begin{tabular}{|c|c|c|c|}
    \hline
    & & \multicolumn{2}{c|}{$\ket{p}$}\\ 
    \hhline{|~|~|-|-|}
    Configuration & $n_r$&5P$_{1/2}$ &6P$_{3/2}$\\
    \hline
    With local & 60 & 0.740 & 0.778 \\     \hhline{|~|-|-|-|}
    addressing & 74 & 0.790 & 0.827 \\    \hline
    With an & 60 & 0.879 & 0.865 \\     \hhline{|~|-|-|-|}
    entanglement zone & 74 & 0.934 & 0.941 \\    \hline
    \end{tabular}
\end{table}

\section{Experimental Description \label{sec:setup}}
\begin{figure}[t]
\centering\includegraphics[width=\linewidth]{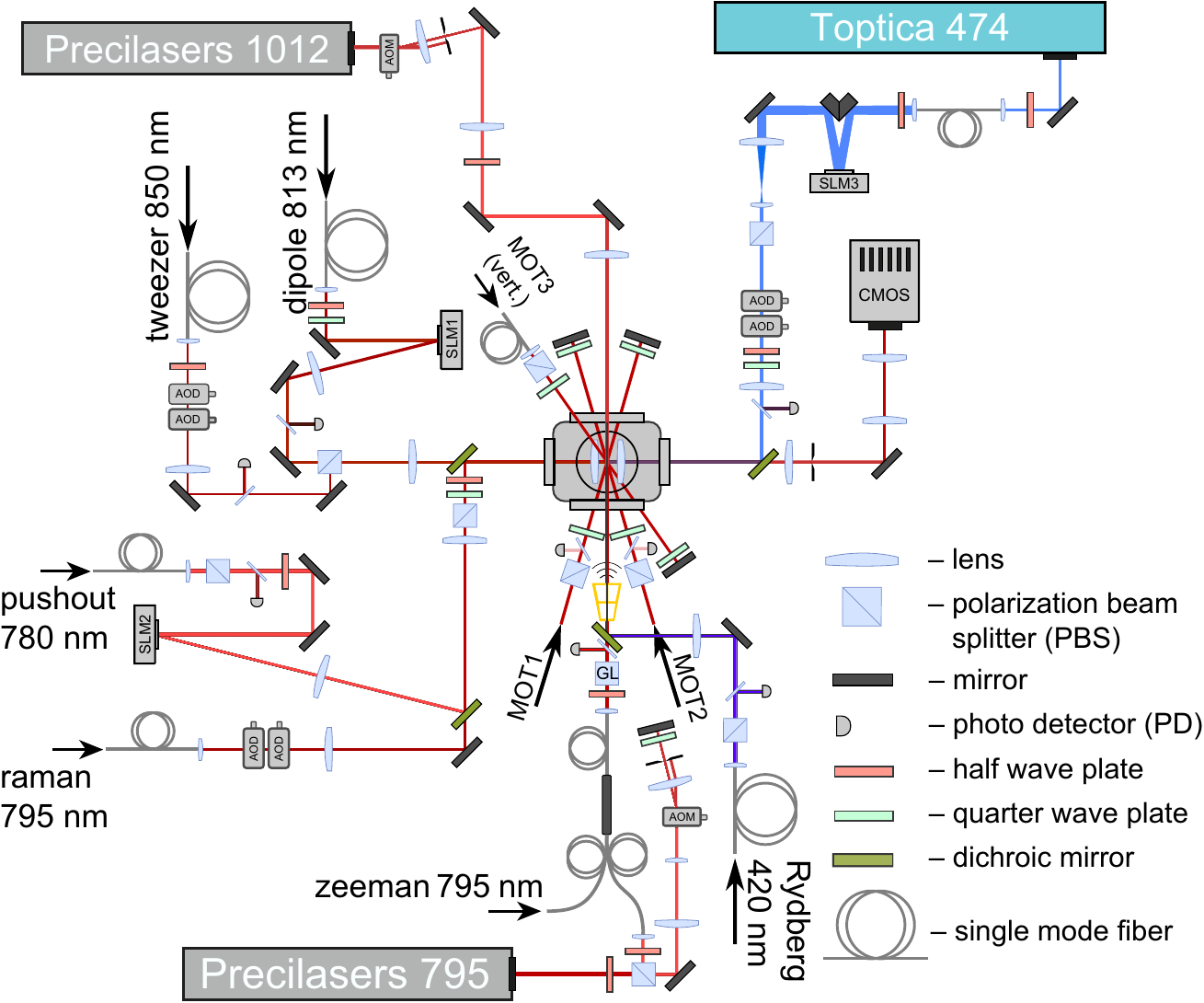}
\caption{Optical scheme of the experimental setup}\label{fig:setup}
\end{figure}

The components of the experimental setup, with the exception of the Rydberg excitation system using the 6P$_{3/2}$ level, are described in detail in Ref.~\cite{Iukhnovets:2026:arxiv}. The optical scheme including the added scheme B is shown in Fig.~\ref{fig:setup}. In the system with 5P$_{1/2}$ (Fig.~\ref{fig:scheme5p}b), the first-stage beam is a broad global-excitation beam with a radius of 0.57~mm (here and below, the radius at the $1/e^2$ intensity level is used to characterize the transverse size of waists $w_0$ and beams $w(z)$). Addressed excitation is provided by the second-stage beam (wavelength 474~nm), which has a flat-top intensity profile at the focus and acts only on the selected pair of atoms. This profile is formed using the algorithm proposed in Ref.~\cite{Iukhnovets:2026:arxiv}, which is based on representing the flat profile as a superposition of the first eight even Hermite--Gaussian modes. The beam is created using a spatial light modulator; hologram formation is based on the approaches described in Refs.~\cite{Davis:1999:38, Bolduc:2013:38}. The wavelengths are chosen so that the Rydberg state 60S$_{1/2}$ is populated. Levels with smaller $n$ are preferable in terms of the system's reduced sensitivity to external electric fields and a smaller blockade radius, but at the same time decreasing $n$ reduces the blockade energy shift at a fixed interatomic distance, and this distance cannot be reduced below 3.6~$\mu$m experimentally.
\begin{figure}[htbp!]
\centering\includegraphics[height=0.8\linewidth]{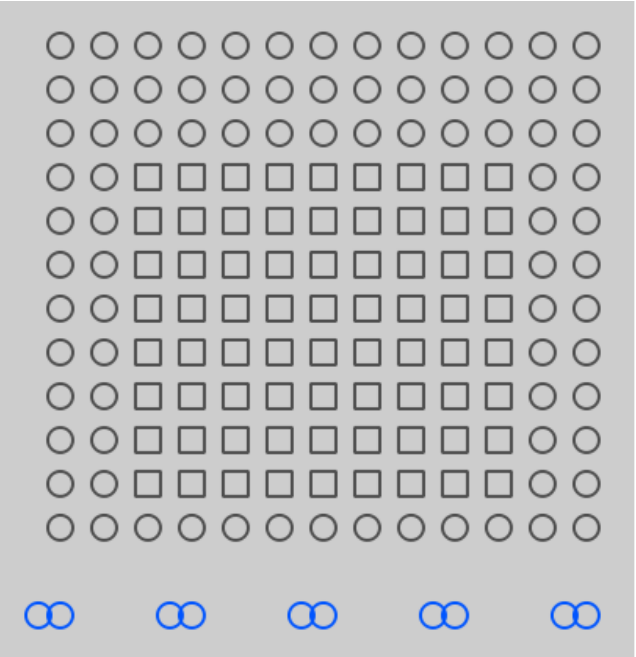}
\caption{Zoned architecture of the atomic array. Black circles show atoms in the reservoir zone, squares show atoms in the computational array. Blue particles in the entanglement zone are shown at the bottom}\label{fig:massive}
\end{figure}

In experiments with the transition through 6P$_{3/2}$, the target qubits are placed in a separate entanglement zone (Fig.~\ref{fig:scheme6p}b). In this scheme, the Rydberg state 74S$_{1/2}$ is excited (the factor limiting $n$ from below is a technical characteristic of the lasers, as already noted in Sec.~\ref{sec:theory}). \par
The spatial configuration of the dipole-trap array as a whole was kept the same as that shown in Ref.~\cite{Iukhnovets:2026:arxiv}; however, the number of particles in the computational array was increased while the total number of atoms (computational array plus reservoir zone) was reduced. Whereas previously the computational array had dimensions $5 \times 10$, in the new implementation it is $8 \times 9$. The size of the reservoir zone was reduced to 84 atoms (Fig.~\ref{fig:massive}). In the configuration corresponding to scheme B, the distance between particles in the array is 7.2~$\mu$m. \par
From the storage zone for computational qubits (the memory zone), atoms are transported to a separate entanglement zone located below 
the computational array at a distance of 14.4~$\mu$m. It is in this zone that two-qubit operations are performed.
\begin{figure*}[htbp!]
\centering
     \includegraphics[width=\textwidth]{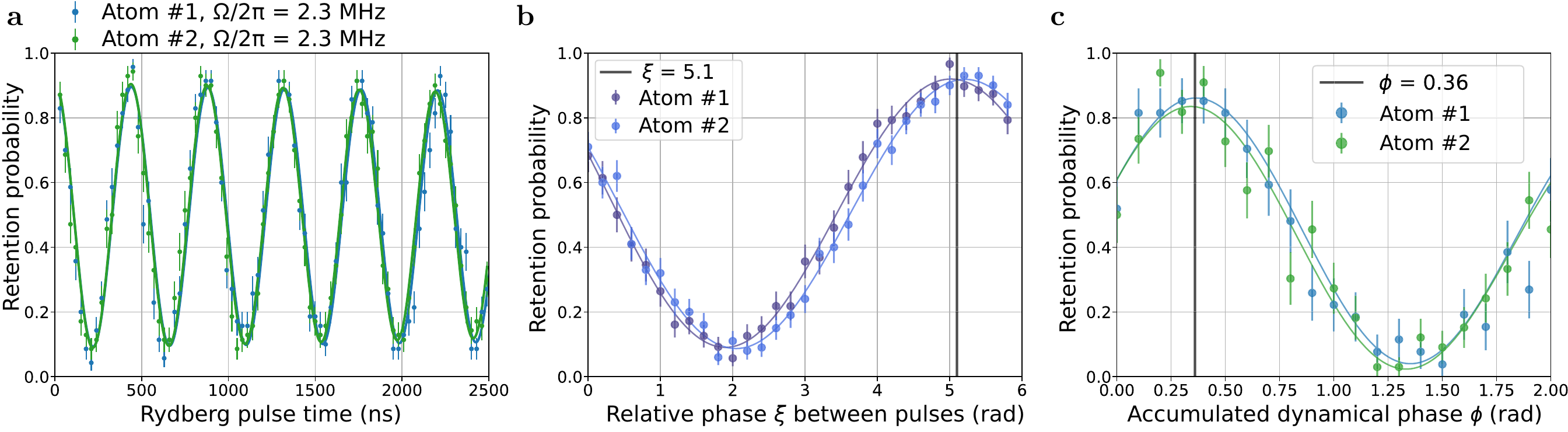}
     \caption{(a) Rabi oscillations under single-particle excitation to a Rydberg state for a pair of atoms in paired sites of the computation region. (b) Dependence of the probability of detecting an atom in the ground state after applying a pair of Rydberg excitation pulses on the relative phase between them. The two curves correspond to separate measurements for each atom of the selected pair. (c) Probability of detecting an atom in the ground state as a function of the compensation phase around the quantization axis. The two curves correspond to separate measurements for each atom of the selected pair} 
     \label{fig:param}
\end{figure*}
\begin{figure}[htbp!]
\centering
     \includegraphics[width=\linewidth]{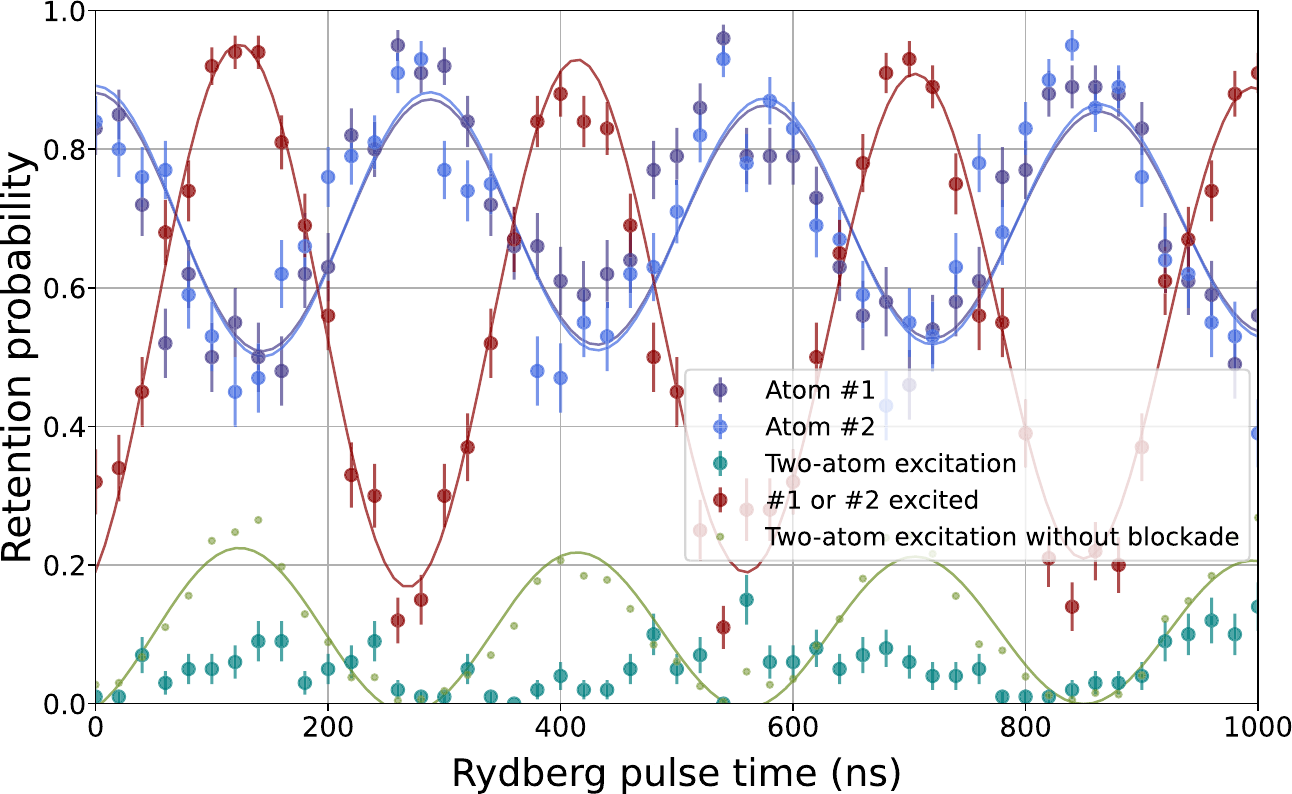}
     \caption{Simultaneous excitation of atoms in paired sites of the computation region to a Rydberg state. The red curve is the probability of exciting one atom of the pair, the green curve is the probability of simultaneous Rydberg excitation of two atoms, the olive curve is the conditional probability of single-particle excitation, and the remaining curves are unconditional probabilities of single-particle excitation}
     \label{fig:blockade}
\end{figure}

The entanglement zone contains 10 atoms organized into five pairs. Atoms within each pair are separated by 3.6~$\mu$m. The distance between pairs is 18~$\mu$m. This configuration makes it possible to treat each pair as a separate two-qubit system while preserving spatial separation between them. \par
Excitation of atoms in the entanglement zone is carried out by two Rydberg beams propagating toward each other in the plane of the trap array (the XY plane in Fig.~\ref{fig:scheme6p}b). Both lasers have a TEM$_{00}$ transverse spatial mode, and their optical paths contain no elements that perform complex phase modulation. The first-stage beam has a wavelength of 420~nm, a power of 3.5 mW, and a waist radius of 15~$\mu$m (the maximum intensity at the waist is 9.62 W/mm$^2$). The second-stage beam $\lambda_2=$1012~nm propagates from the opposite side. It has a power of 6~W and a waist radius of 20~$\mu$m (maximum intensity 9.55~kW/mm$^2$). \par
The coordinates of the waists of both beams along the X axis lie within the region containing the atoms, which has a width along this axis of no more than 100~$\mu$m. Using a profilometer based on gradual opening of the beam by a blade edge \cite{Skinner:1972:5, Landry:2008:phd}, the beam radii were measured and it was confirmed that, within the array zone, the beam radius does not exceed the waist radius by more than 5\%. \par
Stabilization of both lasers in scheme B is carried out using a highly stable four-cavity Fabry--Perot resonator, each cavity of which is bounded by glass surfaces with antireflection coatings optimized for different wavelengths.
The feedback circuits are assembled using a PreciLock radio-frequency modulation and signal-processing device. \par
In both configurations, the CZ gate is implemented according to the Levine--Pichler scheme~\cite{Levine:2019:123}. The 1012~nm beam is switched on in pulses, and during its action two 420~nm radiation pulses of equal duration $\uptau$, separated by the phase $\upxi$, are applied. This phase, as well as the detuning from exact Rydberg resonance $\updelta$, are chosen so that the evolution of both the singly excited state and the state $\ket{\mathrm{W}}=\frac{1}{\sqrt{2}} \left(\ket{r0}+\ket{0r} \right)$, which is excited in the blockade regime, are closed, while the phases accumulated during the evolution differ by $\pi$. \par
For the Rydberg excitation beams with the parameters considered here, the experimentally determined Rabi frequency is $\Omega/2\pi= 2.3$~MHz. To determine it, the duration of the excitation pulse was scanned at exact resonance (Fig.~\ref{fig:param}a). \par
According to Ref.~\cite{Levine:2019:123}, the value satisfying the above requirements is $\updelta=0.377371\Omega$, and the duration of the excitation pulses is (rounded to the nearest multiple of ten)
\begin{equation}
    \uptau = \frac{2\, \pi}{\sqrt{\updelta^2+2 \Omega^2}}=290\, \mathrm{ns}.
    \label{eq:tau}
\end{equation}
After setting these values, we checked for the presence of blockade under simultaneous excitation of atoms in the selected sites (Fig.~\ref{fig:blockade}).
\begin{figure*}[htbp!]
\centering
    \includegraphics[width=\textwidth]{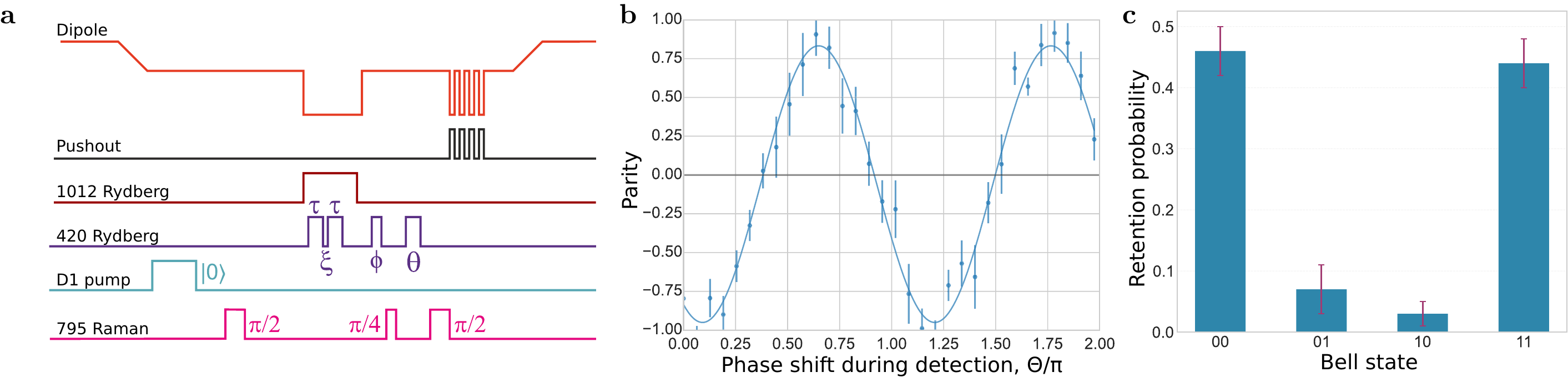}
    \caption{(a) Experimental sequence for measuring the fidelity of the CZ gate. (b) Parity oscillations for the entangled state of a pair of atoms in neighboring traps. (c) Measurement probabilities in the computational basis for the Bell state $\ket{\Phi^+}$} 
    \label{fig:parity}
\end{figure*}
\begin{figure}[htbp!]
\centering
     \includegraphics[width=\linewidth]{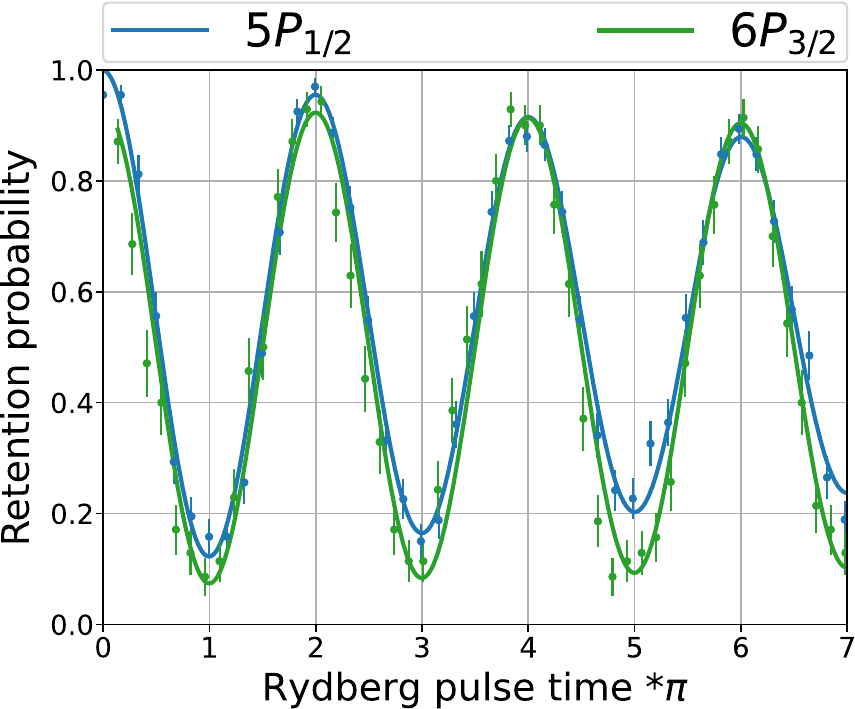}
     \caption{Rabi oscillations of the Rydberg transition using local excitation through the intermediate level 5P$_{1/2}$ (cyan) and global excitation through 6P$_{3/2}$ (green)}
     \label{fig:comparison}
\end{figure}

To determine the phase shift $\upxi$, a pair of pulses with fixed duration $\uptau$ is applied in turn to single atoms in each of the sites (only one of the selected sites is filled). By scanning $\upxi$ between the pulses, the phase value at which an extremum is reached in the resulting dependence (Fig.~\ref{fig:param}b) is determined. This corresponds to closure of the trajectory on the Bloch sphere, that is, after the action of the pair of pulses a single atom returns exactly to the ground state. The phase values must be identical for atoms in neighboring traps. \par
The operation implemented by the tuned pulse pair coincides with CZ up to a global rotation $R_z (\upphi)$, which is determined in the Ramsey scheme (Fig.~\ref{fig:param}c) and compensated by a global pulse of the 420 nm Rydberg excitation laser (the $\upphi$ pulse in Fig.~\ref{fig:parity}a).

\section{Results \label{sec:results}}
For both implemented configurations, Rabi oscillations of the Rydberg excitation were measured. Less strongly damped oscillations were obtained in the implementation of scheme B, despite the larger value of $n$ compared with scheme A (Fig.~\ref{fig:comparison}). This is explained not only by the properties of the transition itself through the 6P$_{3/2}$ level, but also by the features of its experimental implementation: unlike scheme A, this configuration has no addressed excitation of individual atoms, which loosens the stringent requirements on the spatial profile of the excitation radiation. Forming a beam with an almost flat intensity distribution in the region of the addressed atoms while simultaneously suppressing illumination of neighboring sites is one of the most difficult tasks in such experiments. Any imperfection in such a beam inevitably leads to additional inhomogeneities in the excitation parameters and thereby contributes to decoherence. \par
To estimate the fidelity of the CZ gate, a standard method based on parity measurements is used. Figure~\ref{fig:parity}a shows the experimental sequence for such a measurement. A pair of atoms is prepared in the state $\ket{\Phi^+}=\frac{1}{\sqrt{2}}(\ket{00}+\ket{11})$, after which a global rotation $R_Z(\uptheta)\, \otimes \, R_Z(\uptheta)$ is applied to both atoms of the pair and the mean parity value $P=\langle Z \otimes Z \rangle$ is measured as a function of $\uptheta$. The plot of the dependence obtained for scheme B is shown in Fig.~\ref{fig:parity}b. The amplitude of the parity oscillations is equal to twice the off-diagonal element of the density matrix of the resulting entangled state, 2$\rho_{00,11}$, and makes it possible to estimate the fidelity as
\begin{equation}
    F=\langle \Phi^+ | \rho | \Phi^+ \rangle=\frac{1}{2} \left(p_{00}+p_{11} \right) + \rho_{00,11},
\end{equation}
where $p_{00}$ and $p_{11}$ are the populations of the corresponding states. Their values were $p_{00}=0.46 \pm 0.04$ and $p_{11}=0.44 \pm 0.04$ (Fig.~\ref{fig:parity}c). On the basis of the obtained results, the fidelity of the two-qubit CZ operation using scheme B was found to be $F=0.937 \pm 0.05$ with correction for state-preparation-and-measurement (SPAM) error.
\begin{figure}[htbp!]
\centering
     \includegraphics[width=\linewidth]{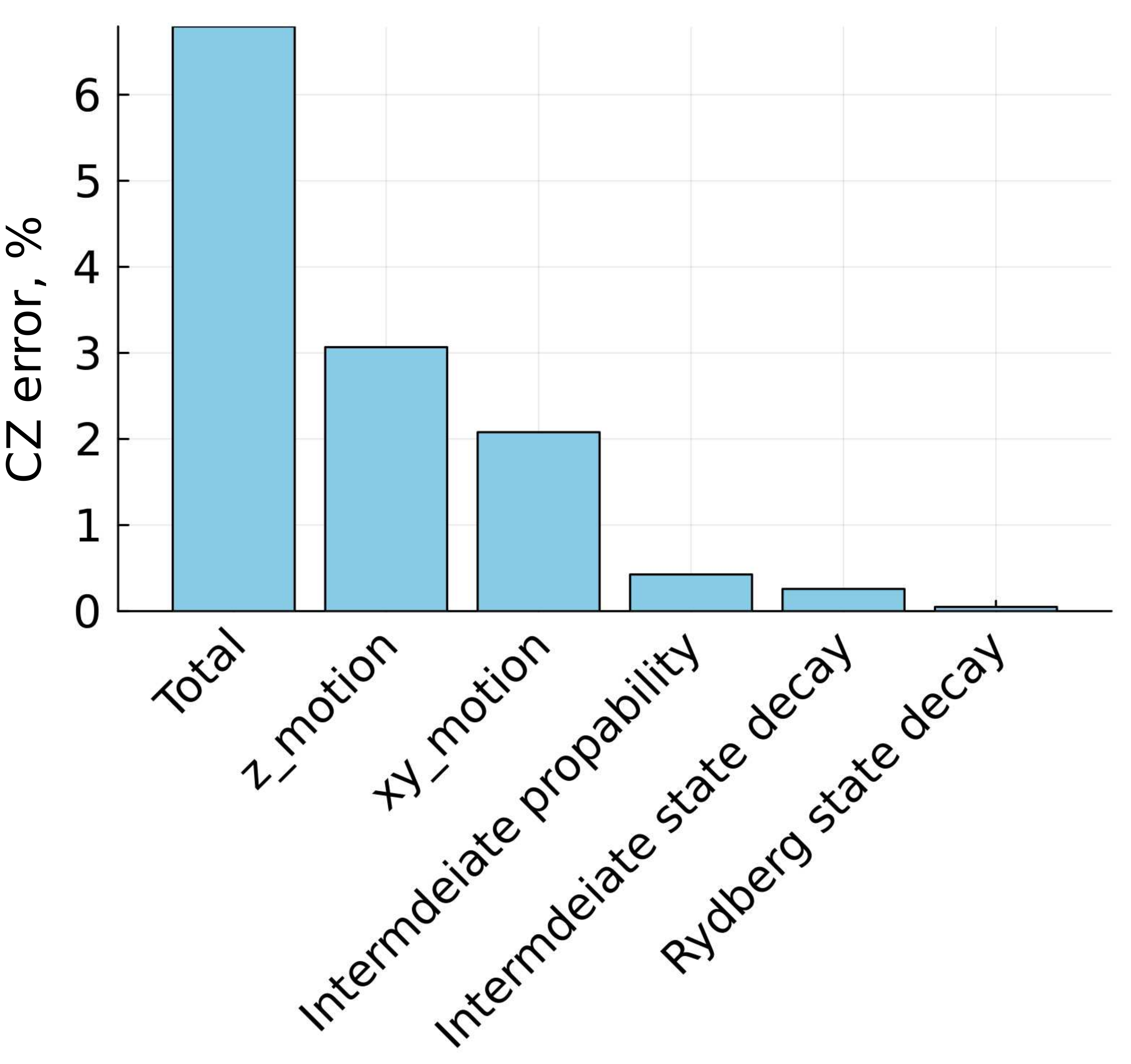}
     \caption{Error budget for implementation of the CZ gate}
     \label{fig:errbudget}
\end{figure}

To determine the SPAM error, the qubit was prepared sequentially in the states $\ket{0}$ and $\ket{1}$, and immediately after each preparation (without performing the gate) the state was measured. As a result of such calibration, a matrix of the type
\begin{equation}
    \mathbf{S}=\begin{pmatrix} P(0|0) & P(0|1) \\ P(1|0) & P(1|1) \end{pmatrix},
\end{equation}
can be formed, where $P(i|j)$ is the probability of registering the state $\ket{i}$ after preparing the atom in the state $\ket{j}$. Using the matrix $\mathbf{S}$, a corrected vector $\vec p^{\,c} = \mathbf{S}^{-1} \vec p$ can be obtained from the vector of measured outcome probabilities $\vec p$ for each qubit. \par
For the two target qubits, the measured SPAM-error matrices are, respectively,
\begin{eqnarray}
\mathbf{S_{1}}=\begin{pmatrix} 0.922 & 0.022 \\ 0.078 & 0.978 \end{pmatrix};
\nonumber\\
\mathbf{S_{2}}=\begin{pmatrix} 0.868 & 0.024 \\ 0.132 & 0.976 \end{pmatrix}.
\end{eqnarray}

Figure~\ref{fig:errbudget} shows the error budget for implementation of the CZ gate obtained using the model described in Sec.~\ref{sec:modeling}. The main contribution to the imperfection is made by the residual thermal motion of atoms in the dipole traps. In the limit $T\rightarrow 0$, with all other system parameters preserved, the fidelity approaches 99.27\%.
\section{Conclusion}
A setup allowing implementation of Rydberg excitation according to two different schemes has been created and experimentally tested. Scheme B, corresponding to excitation through the intermediate level 6P$_{3/2}$, has been implemented on the experimental prototype described here for the first time. To realize it, a two-zone structure of the atomic array with a dedicated computation zone was created; in this zone, global and local single-qubit operations and parallel two-qubit operations are performed using global Rydberg excitation. The size of the memory zone used to store the atomic register was 72 atoms, corresponding to 72 atomic qubits addressed by moving them to the computation zone. \par
The fidelity of the two-qubit gate was determined in an experiment measuring parity oscillations for the entangled state of a pair of atoms and is 0.937. \par
When scheme B is used, Rabi oscillations of the Rydberg excitation decay substantially more weakly than in the earlier configuration, which indicates high coherence of the Rydberg excitation and an increase in the fidelity of two-qubit Rydberg operations relative to scheme A. This advantage is confirmed by numerical modeling and is associated, first, with the absence of tightly focused addressing beams in the experimental implementation and, second, with the physical features of the excitation scheme itself. Implementing addressing is associated with significant difficulties in forming a beam with a flat intensity profile, whose imperfection is an additional source of decoherence. \par
It should be noted here that there is also a fundamentally different approach to eliminating this problem, proposed in Ref.~\cite{Bezuglov:2025:112}, where it was theoretically shown that under coherent three-photon excitation of Rydberg states the effective Rabi frequency can be made almost independent of the atom position. In the considered scheme 5S$_{1/2}\ \rightarrow$ 5P$_{3/2}\ \rightarrow$ 6S$_{1/2}\ \rightarrow n_r$S, with strong coupling at the second stage and moderate coupling at the first and third stages, the effective three-photon Rabi frequency takes the form $\Omega_1 \Omega_3 / \Omega_2$. If the spatial profile of the product $\Omega_1 \Omega_3$ reproduces that of $\Omega_2$ up to a constant factor, the ratio becomes position-independent, and the effective excitation Rabi frequency ceases to depend on the atom coordinate even within tightly focused beams. Thus, the inhomogeneity of excitation parameters that limits coherence can be suppressed not by improving the profile of an individual beam, but by choosing the Rabi frequencies of the individual stages of multiphoton excitation. \par
In the near term, development of the atomic quantum-computer prototype and further improvement of the quantum-computation system are expected to proceed primarily on the basis of scheme B, that is, using the transition through the 6P$_{3/2}$ level. At the same time, improved tight focusing, the three-photon excitation scheme mentioned above, and integration of high-aperture optics into the experiment, in particular objectives with $\mathrm{NA}\geq 0.6$, may open new possibilities for scheme A with local addressing.\par
{\bf Funding.} This work was supported by Rosatom as part of the ``Quantum Computing'' Roadmap under contract No.~868/1653-D dated August 21, 2025.

\end{document}